\documentclass[12pt]{article}

\usepackage[utf8x]{inputenc}
\usepackage{graphicx}
\usepackage{amsmath}
\usepackage{amssymb}
\usepackage{setspace}
\newenvironment{sciabstract}{%
\begin{quote} \bf}
{\end{quote}}

\newcounter{lastnote}

\title{Fly out-smarts man}
\author{ Ruedi Stoop$^1$, Patrick N\"uesch$^1$, Ralph Lukas Stoop$^1$ \& Leonid Bunimovich$^2$\\
\\
\normalsize{$^{1}$Institute of Neuroinformatics, University and ETH of Zurich, CH-8057 Zurich}\\
\normalsize{$^{2}$School of Mathematics, Georgia Institute of Technology, Atlanta, GA 30332-0160 USA}\\
\\
\normalsize{$^\ast$To whom correspondence should be addressed; E-mail:  ruedi@ini.phys.ethz.ch.}
}

\date{}

\begin{document}
\baselineskip24pt
\maketitle


\doublespacing
\begin{sciabstract}
{ Precopulatory courtship is a high-cost, non-well understood animal world mystery. {\it Drosophila}'s (={\it D.'s}) precopulatory courtship not only shows marked structural similarities with mammalian courtship, but also with human spoken language. This suggests the study of purpose, modalities and in particular of the power of this language and to compare it to human language. Following a mathematical symbolic dynamics approach, we translate courtship videos of {\it D.}'s body language into a formal language. This approach made it possible to show that {\it D.} may use its body language to express individual information - information that may be important for evolutionary optimization, on top of the sexual group membership. Here, we use Chomsky's hierarchical language classification to characterize the power of {\it D.}'s body language, and then compare it with the power of languages spoken by humans. We find that from a formal language point of view, {\it D.}'s body language is at least as powerful as the languages spoken by humans. From this we conclude that human intellect cannot be the direct consequence of the formal grammar complexity of human language.}
\end{sciabstract}

\normalfont

\section*{Introduction}
Over the centuries, the evolution of human language has been the subject of controversial discussions among philosophers, linguists and biologists. Yet, a consensus on what causes language to evolve and what are the effects of this on society has not been achieved. Traditionally, language was thought of as a strictly culturally transmitted phenomenon, with few or no biological ties at all. In the second half of the 20th century, under Chomsky's influence who considered that language is located in the brain and therefore is subject to biological conditions \cite{ref3,ref4}, this view started to change. The discussion arose from what the driving force of the evolution of language could be. An important observation is that language - as is any complex ability of humans or animals - is the result of natural selection \cite{ref5}. Chomsky and his school remained, however, skeptical about approaches that saw this as the only driving force. They suggested that language grammar may have emerged as a side-effect of the reorganization of the brain needed for coping with its growing size during evolution towards the modern Homo sapiens \cite{ref6,ref7}. In order to study the evolution of language and to determine its driving forces, a classification of languages accounting for the changes undergone would be helpful. To capture the grammatical complexity aspect of languages, Chomsky and Schützenberger \cite{ref8} proposed a hierarchical classification scheme, comprising grammars of increasing grammatical complexities: type t-3 (
left regular grammar) $\subset$ type t-2 (context free grammar) $\subset$ type t-1 (context sensitive grammar) $\subset$ type t-0 (Turing machine). This classification has proven extremely useful in different fields of comparative sciences. It has been used e.g. to compare spoken human languages, for distinguishing compiler languages, as a basis for the theory of automata, and for classifying dynamical systems.  A natural question is whether the more advanced
organizational forms are generally equipped with more complex language structures?  Our study will answer a more specific - but similarly fundamental - question: Do more complex organizations (society, intelligence,..) require a language representation of increased grammatical complexity? 

To answer this question, we compare the grammatical complexity of human languages (which are known to fall mostly into Chomsky hierarchy type t-2 \cite{ref9,ref10}) with experimental data from the precopulatory courtship body language of the fruit fly {\it Drosophila}. 
In the animal world, courtship ranges from simple rituals to complex communication-like behaviors. Despite its high cost for the animal (energy- and death toll-wise), the origins and purpose of courtship are still not well understood. A natural hypothesis is that courtship is an evolutionary optimization mechanism that a species may or may not take advantage of. Living in a simple and evolutionarily fast environment, {\it D.} provides a well-suited testing case. Until recently, investigations on this nature of {\it D.}'s courtship were hampered by the lack of a conceptual framework able to address this question. Behavior is characterized by rituals that consist of well-chosen sequences of individual actions. Since it is in the nature of these rituals that they need to be repeated if required, we characterize behavior by sequences of indecomposable closed cycles of indecomposable individual actions, so-called irreducible cycles of irreducible acts \cite{ref1,ref2}. This approach is also motivated by the theory of complex dynamical systems, where it has been shown that such systems can be reduced to a minimal set of closed sequences of actions (there called 'irreducible closed orbits'). From this set the system can systematically be approximated by combining ever more of these sequences, starting with the shortest ones (for detailed references cf. Ref. \cite{ref2}). Using a decomposition of such data into irreducible cycles, it has been found that with high confidence during {\it D.}'s precopulatory courtship, individual information is transmitted to the prospective partner, i.e. a real communication with essential information exchange takes place between the partners \cite{ref1,ref2} (supplementary material's Fig. 2).

The focus of the present work is on the power of the grammar that underlies the generation of the courtship language. To the best of our knowledge we use here for the first time Chomsky's classification scheme to characterize courtship and animal body language. Although the question by what grammar a given experimental data was generated is in its narrower sense undecidable \cite{ref11}, we are able to provide an answer in the statistical sense: Namely, we show that it is very unlikely that {\it D.}'s body language is generated by grammars of complexity lower than those of human languages. For some data we find indications of a type t-1 grammar underlying their generation, which reaches beyond the grammatical complexity of human language. An overview of the experimental and computational procedures is presented in Fig. 1.

The data that we use in this study originates from experiments where the courtship behavior of a pair of fruit flies  is recorded in an observation chamber at fixed environmental conditions of $25\,^{\circ}{\rm C}$ and 75\% humidity. From high-speed camera recordings of 30 frames per second, we isolated 37 fundamental behavioral acts and coded the recordings accordingly  \cite{ref1} (supplementary material's Fig. 1). Fundamental acts are body movements that can freely be combined with each other. Besides pairing single normal females in the immature, mature and mated states with single normal males, additionally fruitless mutant males \cite{ref1} were paired with either mature females or with mature normal males, leading to five types of experiments. Since either of the protagonists gives rise to a time series, ten classes of experimental time series were obtained in this way.
Tagging each fundamental act by an integer number, each camera episode is represented by a string or time series of these symbols. A mature female as the protagonist in the presence of a normal male, e.g., generates in this way a time series as\\
$\omega=\{9, 17, 21, 20, 17, 20, 6, 21, 6, 21, 17, 18, 21, 25, 20, 17, 20, 21, 17, 18, 21, 17, 20, 9, 17, 20, 21$, \\
$20, 21, 17, 21, 17, 18, 21, 17, 21, 20, 24, 17, 18, 20, 21, 17, 21, 20, 17\}$.


\section*{Statistical generative grammar model}

The simplest grammatical model for the putative generation of the experimental time series is a a grammar of type t-3 of the Chomsky hierarchy 
of languages. This model is equivalent to a random walk on the given set of symbols with probabilities given by the symbol frequencies observed in the respective experiments, but with no further restrictions imposed. If {\it D.}'s body language is of low complexity, the observed strings should fit well into the random walk model. From simulating the random walk based on the observed symbol probabilities of each experiment, we obtained from each experimental file a set of surrogate files to compare with (see Fig. 1A; throughout our investigations, we use $N_{sim}=100$ simulated random walks). For the comparison, a figure of merit is used.
Every time series $\omega=\{x_0,x_1,...,x_L\}$ is characterized by products along the string of the probabilities $P_{in}(x)$ - measuring that a random walk starting at $x_0$ ends at point $x$ - with $P_{out}(x)$ measuring the probability that a random walk starting at $x$ reaches point $x_L$. 

For the unrestricted random walk, these probabilities are
\begin{eqnarray*}
P_{in}(x) & = & \frac{n!}{n_1!\cdot ... \cdot n_{n_{symb}}!}\cdot p_1^{n_1}\cdot ..\cdot p_n^{n_{n_{symb}}},\\
P_{out}(x) & = & \frac{(N-n)!\,\,\,\,\,\cdot \,\,\,\,\,p_1^{(N_1-n_1)}\cdot .. \cdot p_n^{(N_{n_{symb}}-n_{n_{symb}})}}{(N_1-n_1)!\cdot ...\cdot (N_{n_{symb}})!},
\end{eqnarray*}
where $n$ is the number of steps needed to reach point $x$, producing $n_j$ repetitions of the symbol tagged with index $j$.

The entropy $H_{through}$ associated with a string realization is based on the local walk-through probability $P_{through} := P_{in}\cdot P_{out}$, evaluated along the string, as
$$
H_{through}(\omega):=-\frac{\log(P_{through}(\omega))}{L}=-\frac{1}{L}\sum_{i=1}^{L}\log(P_{through}(x_i))=:\frac{1}{L}\sum_{i=1}^{L}H_{through}(x_i),
$$
with $x_i=(n_1^i,n_2^i,...,n_{n_{symb}}^i)$ the coordinate of point $x_i\in\omega$ in the symbol space. In the figures, $H_{through}$ will always be displayed as $H$, see Fig. 1B. 

%

\section*{Courtship language classification}

We evaluated $H_{through}(\omega)$ for each experiment and for the corresponding random walks. For the latter, we also determined the mean values and the standard deviations, see Fig. 1C. Whereas the t-3 model generates strings with similar $H_{through}(x)$ characteristics for approximately one third of the experimental data  for the remaining two thirds, this description fails. A t-3 example is given in Fig. 1B, left panel; an example where the t-3 model fails is shown in Fig. 1B, right panel. In the latter cases, the experimental $H_{through}(x)$ dramatically differs from those obtained for the t-3 model: The experiment's clear peak around position 170 is very unlikely to be reproduced by a simple random walk. The pyramid-like shape with its clear maximum of $H_{through}$ suggests that in the data, an eminent change has occurred in the way of how symbols are chosen from the alphabet. 

To proceed with those experiments that do not fit into a t-3 model, we apply a recursive approach ('t-3, t-2, t-1 model'). We split a string at the point of maximum $H_{through}(x)$, and model the partial strings $\omega_{1}, \,\,\omega_{2}$ separately by corresponding random walks. Strings of the form $\omega=\omega_1\omega_2$ are generated from a t-2 (i.e. context-free) grammar, since a word $\omega=a^n b^n$, $n\in \mathbb{N}$ cannot be created by a t-3 grammar. t-2 grammars reproduce the characteristics of five of our experiments, they remain, however, to be inappropriate for about half of the data. The obvious solution then is to expand the latter into ever more partial walks. 
Technically, for each file we simulate a set of $N_{sim}$ random walks. On this set, we calculate $H_{through}(\omega)$, their average and their standard deviation. If the original file's  $H_{through}(\omega)$ falls within a standard deviation from the computed average, the random walk describes the string well
and the string is considered to be t-3. Otherwise, by splitting the string $\omega$ at the maximum of $H_{through}(x)$, we obtain $\omega_1$ and $\omega_2$. For these partial strings, random walks are then performed separately, and compared to the original data. If they are close enough in the above sense, we consider the string to be t-2. Otherwise we proceed recursively, which implies context-sensitivity \cite{ref12} and therefore a t-1 grammar.
An example of our iterative t-3, t-2, t-1 procedure is given in Fig. 2A.
If we compare $H_{through}$ of surrogate walks generated according to need by t-3, t-2, and t-1 constructions (green points in Fig. 2B), we see that the obtained values are hardly distinguishable from the experimental data.

A natural question is whether the obtained results could in a simpler way be generated by a sequence of type t-3 grammars.  In order to
investigate this possibility, we checked the abundance of irreducible cycles that 
 provide our mathematical basis for capturing behavior \cite{ref1,ref2}. For a succession of type t-3 grammars, their number should not differ in an essential way from the number obtained by simple type t-3 random walks. We observed, however, a massive increase of the irreducible closed cycles from files that we classified as type-2 or type-1, as is exhibited in Fig. 3A. 
This result corroborates the expectation that an increase of the number of closed cycles could serve as the hallmark of higher grammars \cite{ref2} and provides a further argument for the abundant use of higher grammars in {\it D.}'s body language.

\section*{Conclusion and discussion}

The comparison between the behavior of all observed female flies and all observed normal males uncovers that whereas female flies tend to use type t-3 or t-2 grammars, normal males prefer type t-1 (Fig. 3B). This provides  a novel insight into the role of the courtship protagonists depending on their sexual group from the grammatical perspective.
More fundamentally, we stress that in view of the presented results, {\it D.}'s precopulatory body language is not the result of the simplest grammar type t-3 (i.e., a random walk on states of a finite automaton). 
There is a general agreement that natural human languages fall mostly into the type t-2 of Chomsky's characterization (with among the European languages the Swiss-German and the Dutch showing the highest degree of grammatical complexity \cite{ref10}). On the basis of our analysis one can safely say that the {\it D.}'s body language is of no lesser grammatical complexity than the spoken language of humans.

From our findings one also has to conclude that aspects of spoken language that we often take as  given are not reflected in the grammatical complexity. In particular it is not possible to conclude from  language complexity the developmental level / intelligence of an organism. More complex worlds seem not to require more complex grammars. 

The supremacy of human intellect can thus not be founded in the formal grammatical complexity of the language being used. It emerges, rather surprisingly, that lower level species have recursive elements too (recursion is often the key argument for distinguishing between t-3 and t-2 grammars, see Ref. \cite{ref18}). It appears,  however, that only humans have acquired a kind of awareness of theses structures and have learnt to purposefully use them. \\
\\
This work was supported by the Swiss National Science Foundation SNF grant 200021-122276 to R.S.


\vspace{-0cm}
 
\newpage

\begin{thebibliography}{}
\bibitem{ref11}  The 'string problem' is undecidable, see Ref. \cite{ref12}.
\end{thebibliography}

\bibliographystyle{nat} 
{}
\newpage

\begin{figure}
\begin{center}
\includegraphics[width=0.85\textwidth]{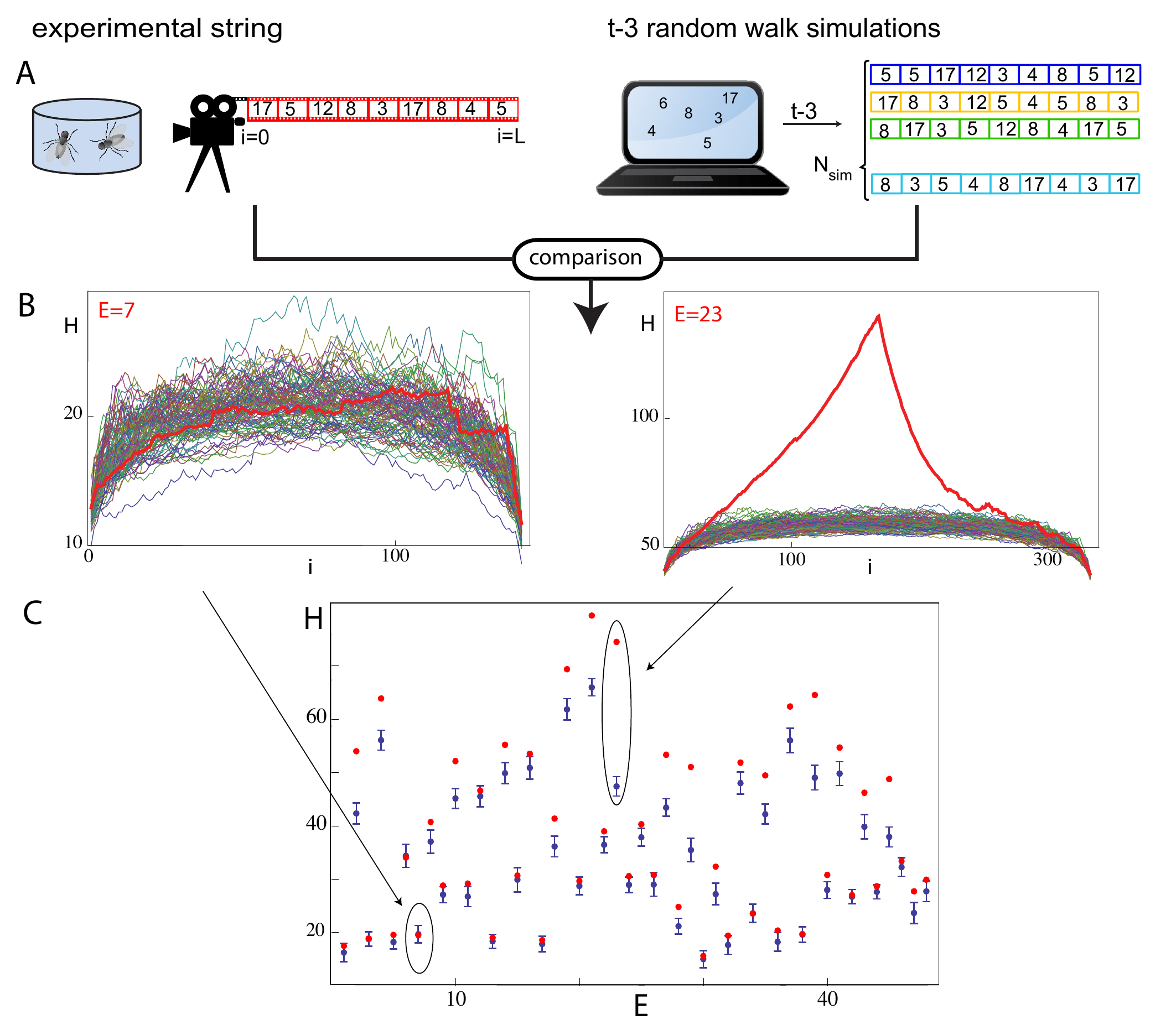}
\caption{\doublespacing A) We compare the experimental symbol strings with strings of equal length generated from a t-3 random walk based on the observed symbol probabilities.
B) For each string (observed and simulated), $H_{through}(x)$ is calculated.  Thick red lines: experiments, thin lines: t-3 random walks. C) $H_{through}(\omega)$
calculated across the data set wraps up the results: Red dots: Experiments; blue dots: Mean values of $N_{sim}=100$ t-3 random walks; bars: one standard deviation. For two thirds of all files, the t-3 model fails.}
\label{fig:Hneu}
\end{center}
\end{figure}

\begin{figure}
\begin{center}
\includegraphics[width=0.85\textwidth]{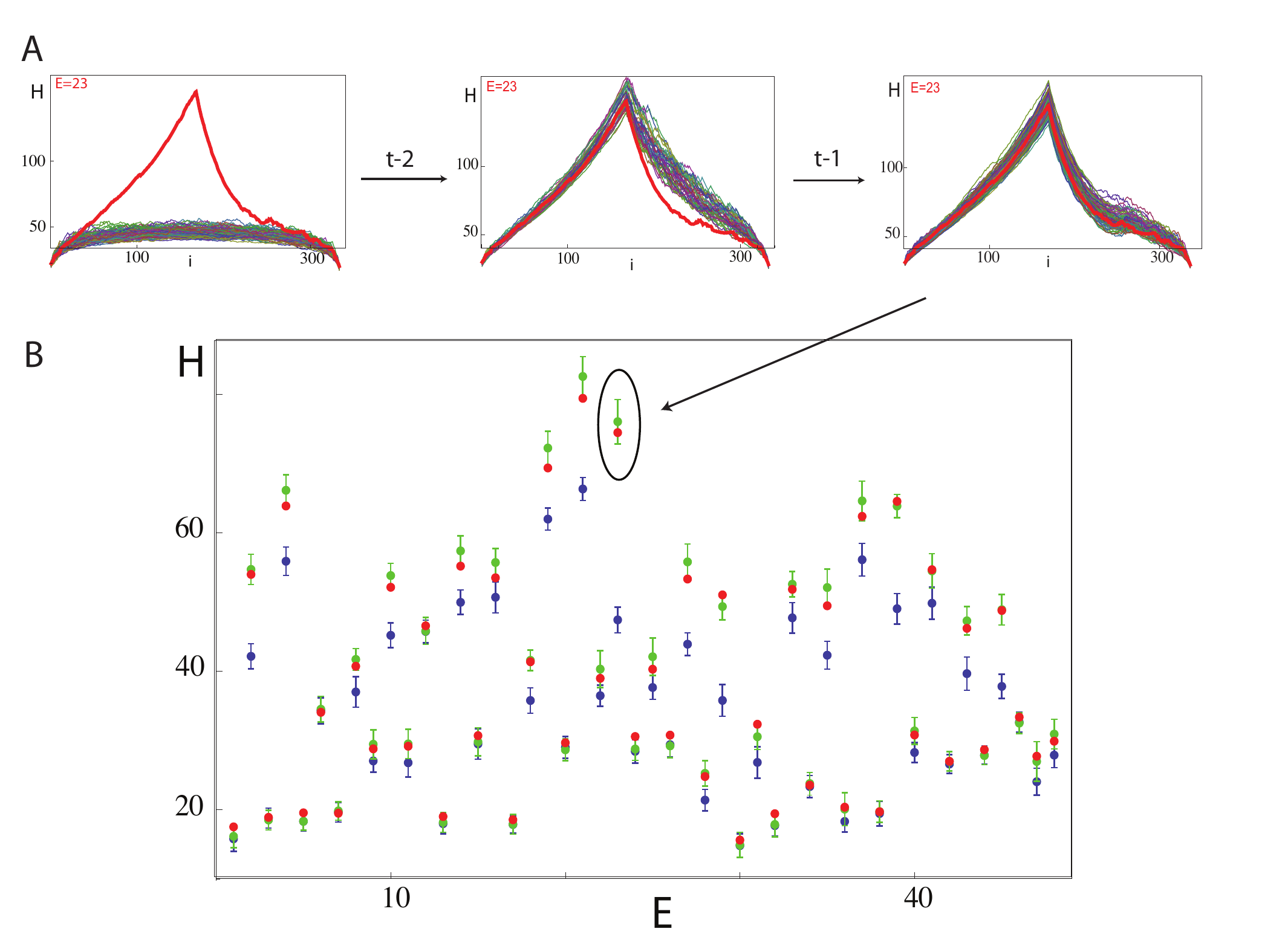}
\caption{\doublespacing A) Experiment E=23 requires a t-1 random walk. 
B) $H_{through}(\omega)$ for each experiment and its surrogate set. Red dots: experiments. Blue: dots: mean values of $N_{sim}=100$ t-3 random walks; bars: one standard deviation. Green: mean values of $N_{sim}=100$ t-3, t-2, t-1 random walks; bars: one standard deviation. Blue dots and bars of t-3 files are obscured by red and green dots and green bars. One can clearly see that the green dots approximate the experimental red dots much better than the blue dots.}
\label{fig:Hneu}
\end{center}
\end{figure}

\begin{figure}
\begin{center}
\includegraphics[width=0.8\textwidth]{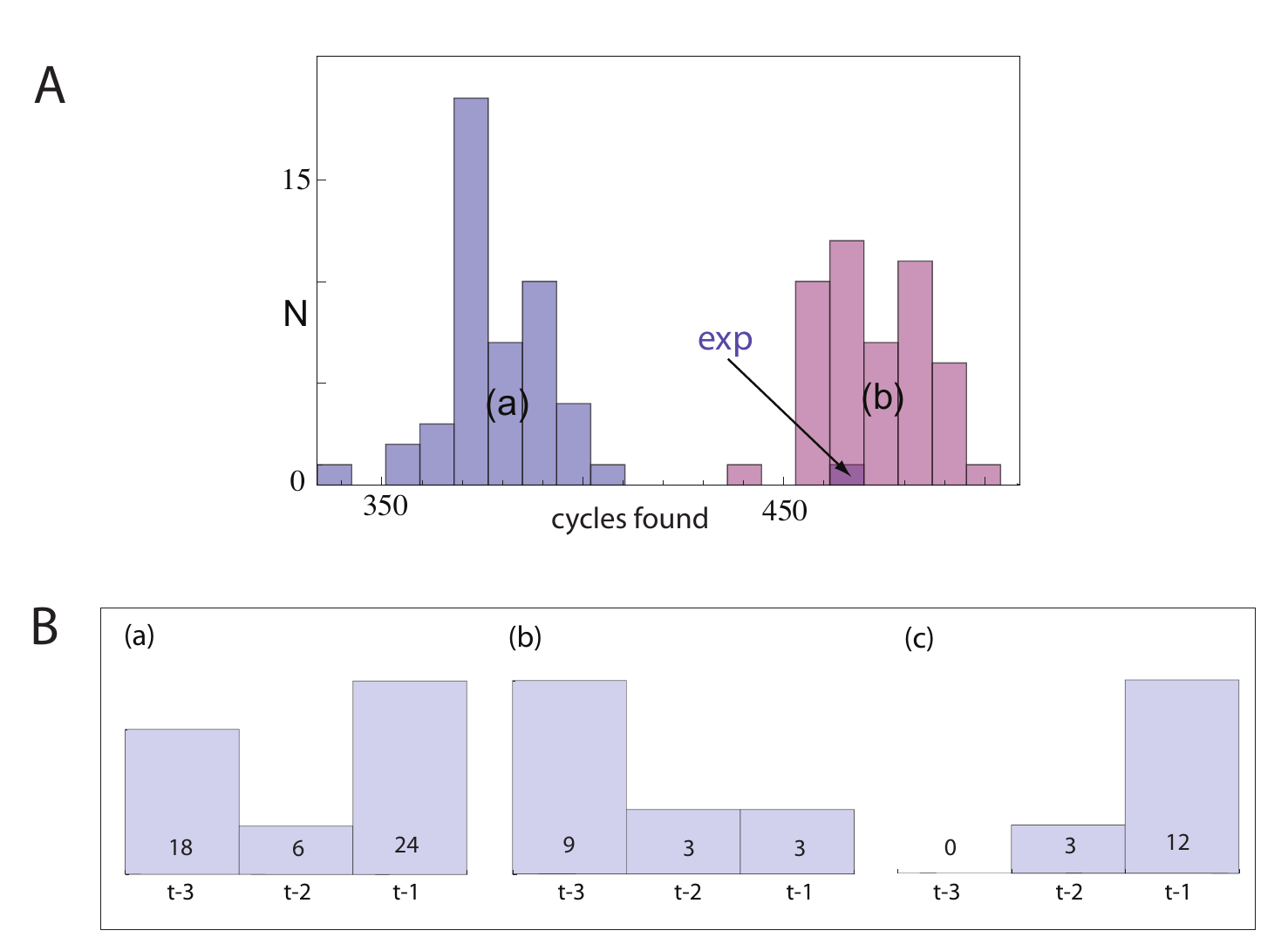}
\caption{\doublespacing A) Histogram of the cumulative number of closed cycles found in a) t-3 random walks,  b) in t-3, t-2, t-1 random walks according to the data's classification, across all data files. The experimental data ("exp") with 468 cycles fits well only into the t-3, t-2, t-1 model. Histograms are based on 100 simulations for all experimental files \cite{ref13}. B) Distribution of t-3, t-2, t-1 classes (with absolute numbers indicated): a) across all experiments, b) across all experiments with females, c) across all experiments with normal males. }
\label{fig:grammar}
\end{center}
\end{figure}

\end{document}